\documentclass[reprint,aps,prl,showpacs]{revtex4-1}
\usepackage{graphicx,epsfig}
\usepackage{epstopdf}% allows eps figures in pdflatex
\usepackage{bm}% allows bold math
\begin{document}

\title{Simultaneous Measurements of the Torsional Oscillator Anomaly and Thermal Conductivity in Solid $^4$He}
\author{D.\,E. Zmeev and A.\,I. Golov}
\affiliation{School of Physics and Astronomy, The University of Manchester, Manchester M13 9PL, UK}
\date{\today}

\begin{abstract}
In these torsional oscillator experiments the samples of solid $^4$He were characterized by measuring their thermal conductivity. Polycrystalline samples of helium of either high isotopic purity or natural concentration of $^3$He were grown in an annular container by the blocked-capillary method and investigated before and after annealing. No correlation has been found between the magnitude of the low-temperature shift of the torsional oscillator frequency and the amount of crystalline defects as measured by the thermal conductivity. In samples with the natural $^3$He concentration a substantial excess thermal conductivity over the usual $T^3$ dependence was observed below 120\,mK.
\end{abstract}


\maketitle

Solid helium is the best example of a {\it quantum crystal}, in which substantial zero-point motion makes atoms and crystalline defects highly mobile even at very low temperatures. Kim and Chan observed anomalies in its response to rotational acceleration \cite{Chan2004}:  
%The latter -- that the resonant frequency of a torsional oscillator (TO) filled with solid $^4$He increases at temperatures below $\sim 100$\,mK -- has generated a stream of publications bearing on various properties of this material. Several groups confirmed the effect, although often of different magnitude and at different temperatures depending on the cell geometry, rate of helium solidification and concentartion of $^3$He impurities, $x_3$ \cite{Reppy06}. 
the resonant frequency of a torsional oscillator (TO) filled with hcp $^4$He increases at temperatures below $\sim 100$\,mK -- suggesting that solid helium no longer accelerates as a rigid rotator, but that a fraction of its mass in effect detaches from the oscillatory motion. This frequency shift disappears when the TO is driven at high amplitudes. Such a shift would have been a sign of the much sought after ``supersolid'', a substance combining crystalline order with superfluidity \cite{Andreev69, Leggett70}. 
The fraction of the effectively detached mass is usually within 0.1\%--1\% but varies for different containers and growth methods.
%Rapid quench of liquid helium, confined to thin annulus, into solid phase seems to result in the highest TO frequency shifts \cite{Chan2004,Reppy07}. 
Annealing of polycrystalline samples often results in the reduction of the frequency shift \cite{Reppy06,Chan2007} but sometimes in its increase \cite{Chan2007}, while single crystals show no change \cite{Chan2007}. 
%Rittner and Reppy \cite{Reppy07} showed that rapid quench of liquid helium, confined to thin annulus, into solid phase results in the highest frequency shifts. Upon subsequent annealing, the frequency shift can be reduced \cite{Reppy07, Reppy10, Chan2007}, left unchanged \cite{Chan2007} or even increased \cite{Chan2007} -- depending on the way the sample is prepared. 
The concentration of $^3$He impurities, $x_3$, in the initial liquid $^4$He strongly affects the characteristic temperature $T_0$ at which the TO frequency is half-way between its high- and low-temperature values: $T_0 \sim 30$\,mK for $x_3 = 10^{-9}$ and increases to $\sim 300$\,mK for $x_3 = 10^{-5}$ \cite{Chan08}. All these observations suggest that the phenomenon is influenced by crystalline defects interacting with impurities. Theoretical explanations have been proposed -- ranging from superfluidity along dislocations and grain boundaries \cite{SuperfluidTheory} to peculiar dynamics of glassy helium \cite{Balatsky} and vibrating dislocations \cite{Iwasa10}. However, no complementary characterization of the density and type of relevant defects has been performed so far. 

Measurements of the phonon thermal conductivity $\kappa$ in solid $^4$He at $T<300$\,mK can probe the quality of the sample.
% since the phonon mean free path (m.\,f.\,p.), $\ell$, is limited by phonon scattering off either crystal defects or container walls \cite{Levchenko}. 
Armstrong {\it et al.} \cite{Armstrong79} studied cylindrical samples of polycrystalline hcp $^4$He and showed that in the defect-scattering regime the phonon mean free path (m.\,f.\,p.) $\ell$ was 0.1--1\,mm, but can be doubled after annealing. 
The goal of our experiments was to see if there is a relation between the value of $\ell$ and the magnitude of the TO frequency shift in different samples of solid $^4$He.

The kinetic-theory expression gives \cite{Ziman}
\begin{equation}
	\kappa = \frac{1}{3}C_{\mathrm{v}}{\bar v}\ell ,
	\label{eq:thcond}
\end{equation}
where the phonon heat capacity per unit volume is 
\begin{equation}
	C_{\mathrm{v}} = \frac{12\pi^4N_{\rm A}k_{\rm B}}{5\hbar^3V_{\rm m}\Theta_{\rm D}^3}T^3,
	\label{eq:Cv}
\end{equation}
and the phonon velocity, averaged over all branches and crystal orientations (for a polycrystal), suitably approximated by the Debye velocity is 
  \begin{equation}
	{\bar v} = \frac{<v^{-2}>}{<v^{-3}>} \approx \frac{k_{\rm B}\Theta_{\rm D}}{2\pi\hbar}\left(\frac{4 \pi V_{\rm m}}{9N_{\rm A}} \right)^{1/3}. 
	\label{eq:v}
\end{equation} 
The heat capacity of hcp $^4$He follows the $C_{\rm v} \propto T^3$ dependence of Eq.\,2 at temperatures $T<500$\,mK, albeit with occasional sample-dependent deviations below $\sim 100$\,mK \cite{Chan2009}. In essence, measuring $\kappa(T)$ is an acoustic probing of a solid by thermal (predominantly transverse) phonons of frequency $f \sim h^{-1}k_{\rm B}T \sim 2$\,GHz and wavelength $\lambda \sim h v_{\rm t} (k_{\rm B}T)^{-1} \sim 0.2$\,$\mu$m (evaluated for $T=100$\,mK). 
Here $v_{\rm t} \propto c_{44}^{1/2}$ is the velocity of the transverse sound. The velocities $v_{\rm t}$ and ${\bar v}$ are expected to be independent of temperature: even though the shear modulus $c_{44}$, measured at low frequencies 0.5\,Hz--8\,kHz, was found to decrease above $\sim 100$\,mK \cite{BeamishShearModulus}, this phenomenon is not expected to extend to our GHz frequencies in the temperature range of interest, 80--250\,mK (at least within the models of glass-like dynamics of solid helium \cite{Balatsky} and mobile dislocation segments \cite{Iwasa10}, whose characteristic frequencies at $T\sim 100$\,mK are $\sim 1$\,kHz and $\sim 10$\,MHz). 

In our experiment, helium samples had a shape of cylindrical annulus of mean radius $r=6.63$\,mm, thickness $d=0.30$\,mm and height $h=14$\,mm. Torsional oscillations were investigated about the cylinder axis while the thermal conductivity was measured along the axis. To estimate the phonon m.\,f.\,p. in the boundary-scattering limit, one can think of a long slab of rectangular cross-section of width $\sim 2(2rd)^{1/2}$ and hence aspect ratio $n \sim 2(\frac{2r}{d})^{1/2} = 13.3$. For diffuse scattering off container walls, this  leads to $\ell \approx \frac{3}{4}(\ln2n+\frac{1}{2})d = 0.85$\,mm \cite{Maris1970}, while exact numerical calculations for our annulus yield $\ell = 0.97$\,mm \cite{Maris2011}. 
The cell was composed of two coaxial stainless steel tubes brazed together at the bottom (see inset in Fig.\,1). Their roughness (asperity size) was on scale $\sim 2$\,$\mu$m, and wall thickness of the inner (outer) tube was 0.5\,mm (0.3\,mm). 
 The assumption of diffuse scattering off these walls should hold for phonons of wavelength 0.2\,$\mu$m even for grazing angles   $\sim d/\ell = 0.3$\,rad that mainly contribute in this geometry. 
\begin{figure}[h]
\includegraphics[width=7cm]{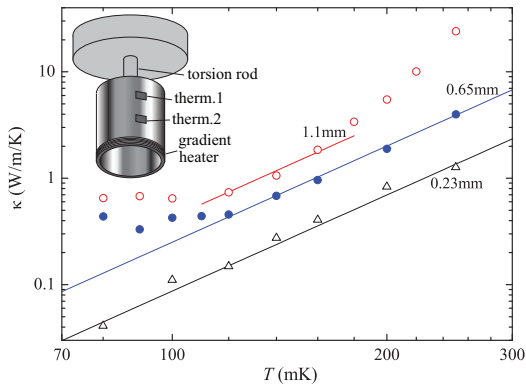}
\caption{Temperature dependence of thermal conductivity for a sample of purified ($\triangle$) and non-purified (as-grown, $\bullet$, and after annealing for 10\, hours at 1.77\,K, $\circ$) $^4$He. The lines indicate $T^3$ dependence and are labelled by the corresponding values of m.\,f.\,p. Inset shows a sketch of the cell.}
\label{fig:dirty_thcond}
\end{figure}

 %The highest experimental m.\,f.\,p. in perfect samples is thus expected to be about 1.0\,mm or perhaps slightly greater for partially specular scattering.
 %, especially at lower temperatures as the wavelength of thermal phonons increases. 
At the top the cell was brazed to a beryllium copper piece consisting of a disk-shaped plate, a torsion rod and a flange. The flange was indium-sealed to a beryllium bronze parallelepiped, capacitively coupled to two flat electrodes for driving and detecting torsional oscillations, with another torsion rod that was attached to the platform with an indium seal. A channel drilled through both torsion rods formed the fill line. Oscillations could be excited with the resonant frequency of 853.6\,Hz and linewidth of 1.05\,mHz (the quality factor $Q=8.1\times10^5$) at $T$=300\,mK. 
%The oscillator consisting of the cell, a dummy bob (the electrode with its flange) and the two torsion rods could thus be excited in two torsional modes (symmetric and antisymmetric) with resonant frequencies equal to 132.5\,Hz and 853.6\,Hz at $T$=300\,mK. The width of the resonant line of the high-frequency mode was 1.05\,mHz (the quality factor $Q=8.1\times10^5$). 
%The low-frequency mode was not used in the present measurements. 
 The platform held a heater and two thermometers (a $^3$He melting curve thermometer and a calibrated germanium resistor) and  was thermally connected to the mixing chamber of a dilution refrigerator through a copper wire, $\sim0.5$\,mm in diameter and $\sim 5$\,cm in length. 
 %For the thermal conductivity measurements the cell had a gradient heater (a wire of 1.5\,k$\Omega$ resistance) wound and glued at the bottom and two RuO$_2$ thick-film resistor thermometers glued with varnish 7\,mm apart. 

Samples were grown from either $^4$He, isotopically purified using heat flush technique \cite{Hendry87} ($x_3<5\times10^{-13}$), or $^4$He with natural isotopic ratio ($x_3\sim3\times10^{-7}$). Liquid helium was pressurized to $P=84$\,bar at 3.5\,K and then cooled down to complete solidification at $P_{\rm m}=51$\,bar and $T_{\rm m}=2.35$\,K (corresponding to $V_\mathrm{m}=19.5$\,cm$^3$ and $\Theta_{\rm D}$=31.5\,K \cite{Driessen86,{Greywall77}}). During the cool-down the fill line was first blocked by solid near the 1\,K pot of the dilution refrigerator, and then the sample in the cell grew at constant volume. The resonant frequency of the TO decreased by 1.6\,Hz upon filling the cell and solidification. To melt and regrow the sample we used either the gradient heater or the platform heater. By varying the heater current and heating time we were able to vary the sample growth time between $\sim 40$\,s and several hours. 

Thermal conductivity was measured by means of running a known current through the 1.5\,k$\Omega$ gradient heater at the bottom of the cell and subsequently measuring the temperature difference established between two RuO$_2$ thick-film resistor thermometers glued with varnish to the side of the cell $z=7$\,mm apart and 3.5\,mm from either end. The end-effect correction to $\kappa$ \cite{Ziman} is small for $\ell \leq 1$\,mm. The heat leak through the thermometer wires was negligible. With homogeneous and isotropic samples of polycrystalline solid helium in the cell, as well as with the empty cell, this arrangement results in a uniform axial distribution of temperature gradient. The temperature of thermometer 1, $T$, was stabilized by applying a computer-controlled power to the heater attached to the platform. At each value of $T$ we used 4--5 different values of the current in the gradient heater (from zero to maximum compatible with the cooling power of the cryostat). Depending on temperature, the process of stabilization took up to 1 hour, after that it could take up to 1 hour to average the readings of the thermometers in the steady state regime. Their temperature difference $\Delta T$ was proportional to the power dissipated in the gradient heater $\dot{Q}$ with an uncertainty of 15\% (at the lowest reported temperature) or less. The value of $\kappa = \frac{z}{2\pi r d} \frac{\dot{Q}}{\Delta T}$ was thus calculated with the absolute accuracy of $\sim 10\%$. The typical values of $\dot{Q}$ ranged from 10\,nW to 1\,$\mu$W, and $\Delta T$ were between 0.1 and 1\,mK, i.\,e. within $10^{-3}$--$10^{-2}T$.

For the measurements of the TO response the temperature has been typically swept from 300\,mK to the lowest achievable temperature, 22\,mK, over 15 hours. On sweeping the temperature twice slower we have not observed any difference; hence, the presented temperature dependences are equilibrium. The TO amplitude was kept sufficiently low, so that 50\% increase of the amplitude of the driving voltage did not result in any decrease of the magnitude of the frequency shift at low temperatures.

The thermal conductivity of the empty cell was found to be proportional to temperature and
consistent with that for stainless steel; it did not exceed 20\% of the total thermal conductivity at the lowest temperatures and was subtracted. We have also measured the temperature dependence of the resonant frequency and linewidth of the empty TO. These were subsequently subtracted from the data obtained with the filled cell. The resonant frequency of the empty cell increased by 0.3\,mHz upon cooling from 300\,mK to 50\,mK, which was much smaller than the typical frequency shifts of the filled cell.

For all 17 samples of purified $^4$He, we observed the $\kappa \propto T^3$ dependence in the whole range of temperatures where we could measure it, 80--250\,mK (see example in Fig.~\ref{fig:dirty_thcond}, $\bigtriangleup$).  
The value of $\ell$, calculated using Eq.\,(1--3), was typically 0.23--0.27\,mm for rapidly grown samples and increased to 0.34--0.40\,mm after annealing. As all these values are smaller than the theoretical upper limit of $\ell \approx 1$\,mm,  phonon scattering off crystalline defects was dominant. Thus the quality of each sample can be characterized by the parameter $\ell$.
%In the boundary-dominating region, if $\kappa$ is measured in-plane in a slab of a large aspect ratio with diffuse scattering at boundaries separated by $d=0.3$\,mm, the m.\,f.\,p. should be \cite{Carruthers1961}
%\begin{equation}
%	 \ell = \frac{3\pi}{8}d = 0.35{\rm \,mm}.
%\label{ell}
%\end{equation}
%Our technique of solidification is expected to produce poor-quality polycrystals. 
%For purified $^4$He, $\ell$ was smaller than the theoretical upper limit of 1\,mm -- both in rapidly grown (0.23--0.30\,mm) and annealed (0.34--0.40\,mm) samples. 
%The fact that $\ell$ was smaller than the theoretical upper limit of 1\,mm indicates that phonon scattering off crystalline defects was dominant, thus the quality of each sample can be characterized by the parameter $\ell$.
%Our values for annealed samples, $\ell =$ 0.35--0.40\,mm, agree with the theoretical upper limit, Eq.\,(\ref{ell}). That in rapidly-grown samples $\ell$ was about a factor of 1.5 shorter indicates that there was considerable additional scattering of phonons by dislocations and grain boundaries.
% but yet with their separation not very much smaller than the slab thickness. 
%Thus the thermal conductivity of purified samples agrees with expectations, so the quality of each sample can be characterized by the parameter $\ell$.

Surprisingly, the thermal conductivity in all 9 samples of non-purified $^4$He was higher than that of purified samples. Moreover, all of them had low-temperature flat parts below $T=120$\,mK (Fig.~\ref{fig:dirty_thcond}, $\bullet$ and $\circ$). The phonon m.\,f.\,p., as inferred from $\kappa \propto T^3$ parts, was 0.60--0.73\,mm for all as-grown samples. Annealing resulted in further increase of $\kappa$ -- by a factor of 1.6--2.0 at $T<150$\,mK and by a progressively larger factor at $T>150$\,mK. The resulting values of $\ell$ (as extracted from $\kappa\propto T^3$ parts at $T \sim 140$\,mK) became 1.1--1.4\,mm -- in fair agreemtnt with the upper limit expected for boundary-scattering regime, $\ell \approx 1$\,mm (perhaps enhanced slightly by specular scattering).  It thus appears that only in annealed non-purified $^4$He phonon scattering was boundary-dominated. 
%The fact that even in annealed samples of purified $^4$He the phonon m.\,f.\,p. was defect-dominated, might suggest that vibrating dislocations scatter thermal phonons more efficiently than those pinned by $^3$He impurities. 

%This result suggests that the excess thermal conductivity is not caused by a moving superfluid potentially responsible for the frequency shift.
 
%Even in as-grown samples, the $\kappa \propto T^3$ regime yields pretty high $\ell \sim 0.4$\,mm but there is also an anomalous plateau in $\kappa(T)$ at 80--120\,mK. After annealing, the TC of non-purified samples increases further, and the $T^3$ behaviour generally changes to an anomalously steep temperature dependence at temperatures 200--250\,mK \cite{Poiseuille}. 
%the m.\,f.\,p. $\ell$ inferred from the measured $\kappa$ is always greater than the theoretical limit of 0.35\,mm, especially in annealed samples. 

\begin{figure}[h]
\includegraphics[width=7cm]{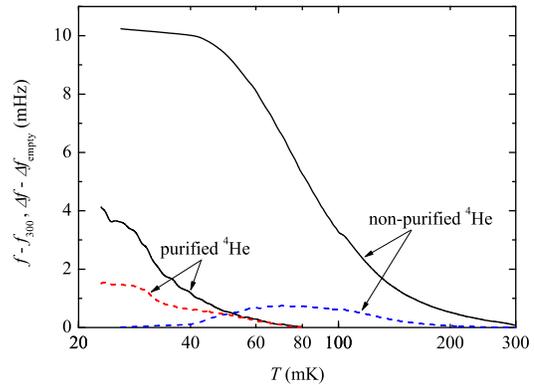}
\caption{TO resonant frequency $f$ (solid lines) and linewidth $\Delta f$ (dashed lines) for samples with different $x_3$.  The value of the resonant frequency at $T=300$\,mK, $f_{300}$, was subtracted. The frequency shift of 1\,mHz corresponds to the effective decoupling of 0.064\% of the mass of solid $^4$He. The change of 1\,mHz in the linewidth corresponds to 1.2$\times10^{-6}$ change in the damping, $Q^{-1}$, of the TO. 
%Thin (blue) lines show the fit of the Iwasa ansatz \cite{Iwasa10} to the non-purified $^4$He data.
}
\label{fig:dirty}
\end{figure}

\begin{figure}[h]
\includegraphics[width=7cm]{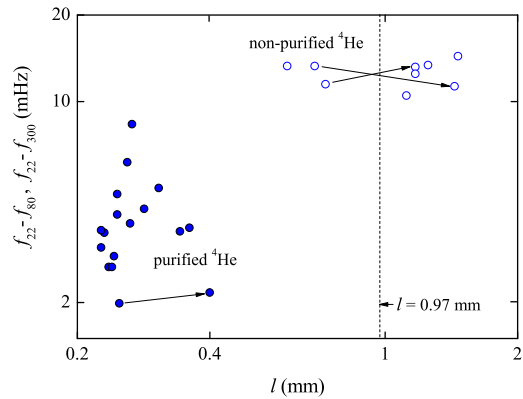}
\caption{The shift in the TO resonant frequency (between $T=80$\,mK and $T=22$\,mK for purified $^4$He, $\bullet$; and between $T=300$\,mK and $T=22$\,mK for non-purified $^4$He, $\circ$) vs. the phonon m.\,f.\,p. Arrows show changes upon annealing at $T=1.77$\,K for 10--13 hours.}
\label{fig:ncrifl}
\end{figure}

Fig.~\ref{fig:dirty} shows examples of the TO resonant frequency $f$ and linewidth $\Delta f$ for samples of purified and non-purified $^4$He. In the latter, the characteristic temperature is $T_0 \approx 80$\,mK. 
%We note that the measured TO characterisitics are well described by the formuli suggested by I.~Iwasa in the framework of his dislocations vibration theory \cite{Iwasa10}. 
In purified $^4$He the cross-over temperature is $T_0 < 30$\,mK \cite{Chan08}, hence we could not reach the saturation of the frequency shift at our lowest attainable temperature $T=22$\,mK. Nevertheless, the resonant frequency shift between 80\,mK and 22\,mK, $f_{22}-f_{80}$, can be used as a measure of the size of the effect under discussion. 
%An example of the TO measurements of a sample grown from non-purified $^4$He is presented in Fig.~\ref{fig:dirty}. 
%Such samples systematically showed a larger frequency shift and smaller linewidth change than in the case of purified $^4$He. 
Using different rates of growing and annealing we were able to obtain samples of purified  $^4$He with $\ell$ ranging from 0.23 to 0.40\,mm and measured the temperature dependence of resonant frequency for all of them. To our surprise, we have not observed any correlation between the magnitude of the frequency shift and $\ell$ in these samples (Fig.~\ref{fig:ncrifl}). Similarly, in samples of non-purified $^4$He, no correlation was found between the TO frequency shift $f_{22}-f_{300}$ and thermal conductivity (characterized by the value of $\ell$ extracted from the $\kappa \propto T^3$ part at $T > 120$\,mK).
%For instance, it would not be surprising if $\ell$ were mainly controlled by the scattering of phonons off grain boundaries while the TO frequency might be more sensitive to the presence of uncorrelated dislocations. 
%Annealing not too close to the melting temperature (as in our case of $T=1.77$\,K = $0.8T_{\rm m}$) then might efficiently remove stress and restructure dislocations into walls but not cause coarsenning of grain structure. 

In different as-grown samples, $f_{22}-f_{80}$ for purified $^4$He were in the range 2--8\,mHz, and $f_{22}-f_{300}$ for non-purified $^4$He were within 10--15\,mHz. Subsequent annealing was not found to change the values of these parameters significantly. For example, annealing at 1.77\,K for 10--13 hours only resulted in changes of the frequency shift and linewidth within $\pm 10$\,\% in both purified and non-purified samples; at the same time $\kappa$ typically increased by 50\%--100\% (Fig.\,3). This is in contrast to a substantial reduction of the frequency shift observed in highly-disordered samples \cite{Reppy06,Chan2007} but in agreement with little effect of annealing on the frequency shift in either already well-annealed polycrystals or single crystals \cite{Chan2007}(albeit containing dislocation walls mobile above $\sim 0.8\,T_m$ \cite{BurnsXray2008}) . We can thus speculate that our samples, even if polycrystalline, might begin with large grains of a size on the order of the gap width, and these grains do not coarsen any further; on the other hand, dislocations that could be efficient scatterers of phonons (especially those freely vibrating in purified $^4$He) can still relax readily.  
%Similarly, in samples of non-purified $^4$He, annealing at 1.77\,K for 10\,hours has resulted in a small {\it increase} in the resonant frequency shift (by 8\% at $T=40$\,mK) and linewidth (by 10\% of the peak value at $T=80$\,mK). 
% (and by a larger factor in natural $^4$He at $T>150$\,mK, Fig.~\ref{fig:dirty_thcond}). 
%Again, no correlation was found between the TO frequency shift and phonon m.\,f.\,p. (as extracted from the $\kappa \propto T^3$ part, if any, at $T> 120$\,mK) in different samples.
The observed absence of correlation between the TO frequency shift and phonon m.\,f.\,p. does indicate that these two parameters are sensitive to different types of disorder. For instance, the former could be more sensitive to grain boundaries while the latter -- to uncorrelated dislocations. 

The upward deviations from $\kappa \propto T^3$ at $T<120$\,mK, observed in non-purified $^4$He, cannot be accounted for by the enhancement of specular scattering as the wavelength of thermal phonons increases -- because the weak temperature dependence $\ell(T)$ observed in this regime (even for phonon m.\,f.\,p. much greater than the asperity size and for slabs of a higher aspect ratio $n \sim 20$ \cite{Eddison1985}) would only lead to a slight reduction in the slope of $\kappa(T)$, and thus can hardly account for the observed plateau-like $\kappa(T)$. Neither can this anomaly be attributed to end effect caused by increased $\ell$ -- as this would {\it decrease} the apparent $\kappa$ (and by different extent for different $\ell$). Last but not least, the presense of crystalline defects limiting m.\,f.\,p. to $\ell < 1$\,mm (e.\,g. $\ell = 0.65$\,mm in Fig\,1, $\bullet$) would prevent grazing-angle phonons from propagating far even if boundary scattering becomes specular. 
%Hence, the thermal conductivity of non-purified samples cannot be explained by the standard phonon model.
%We should hence conclude that the thermal conductivity of non-purified samples cannot be explained by the standard phonon model.
We should hence conclude that there exists an additional channel of heat transport in parallel with Debye phonons. Indeed, an excessive contribution to the specific heat \cite{Chan2009} at similar temperatures have been observed recently, although never yet in the samples of solid helium that can be simultaneously characterized by the TO effect. 
If both effects are caused by the same underlying phenomenon, it seems unlikely that it is superfluidity (that in liquid $^4$He boosts heat transport by means of the counterflow of the normal and superfluid components). This is because 
we found that $\kappa$ is independent of the TO amplitude: with a sample of non-purified $^4$He at $T=90$\,mK with the gradient heater turned on we were increasing the TO drive amplitude by a factor of 50 -- until both the TO frequency shift and excess damping disappeared (the superfluid component would have vanished at this point), but the temperature difference did not change within the experimental accuracy of $\sim 5$\%.
It would be tempting to attribute the observed plateau in thermal conductivity to the gas of $^3$He impuritons at concentration $x_3$ and m.\,f.\,p. of $\sim 1$\,mm. However, even with a generous assumption of the impuriton band of $0.1$\,K, one can only account for $10^{-4}$ of the observed value. %of $\kappa \sim 0.5$\,Wm$^{-1}$K$^{-1}$. 
%For instance, the temperature dependences of our measured $\kappa(T)$ and of the excess specific heat observed by Lin {\it et al.} \cite{Chan2009} in certain samples are very similar. 
%Propagating excitations, different from phonons, were reported \cite{Goodkind2002}, although in $^4$He with fewer $^3$He impurities, $x_3 \sim 10^{-9}$. 

We note that Armstrong {\it et al.} \cite{Armstrong79}, who used $^4$He of apparently natural purity (i.\,e. $x_3 \sim 3\times10^{-7}$), did not observe any deviations of thermal conductivity from $T^3$ dependence down to $T=30$\,mK. It is likely that the reason in the difference from our observations lies in the geometry of the experimental volumes that can affect the types and density of crystal defects: a long cylinder of 3\,mm diameter in \cite{Armstrong79} versus essentially a 0.3\,mm-thick slab in our case. It is plausible that confinement between parallel walls not only allows rapid quench from liquid into the solid phase but also helps to arrest relaxation and annealing of dislocations and grain boundaries. 
%Also, specular scattering, if relevant, is of greater importance in thin-annulus than bulk cylinder geometry. 
% through both pinning and reduction of effective dimensionality.  

The upward deviation from $\kappa \propto T^3$ at high temperatures 160--250\,mK (Fig.\,1, $\circ$), was only observed in non-purified samples after annealing and will require further investigation. Even though it looks similar to $\kappa \propto T^6$ dependence \cite{Armstrong79,Mezhov65} that is often observed in perfect samples at higher temperatures $T\sim 500$\,mK at which frequent normal phonon-phonon scatterings can result in an effectively enhanced $\ell$ (Poiseuille flow), at $T<250$\,mK the phonon gas is too rarified for hydrodynamic behaviour.

%The fact that even in annealed samples of purified $^4$He the phonon m.\,f.\,p. was defect-dominated, might suggest that vibrating dislocations scatter thermal phonons more efficiently than those pinned by $^3$He impurities. 

%What could be the origin of the excess thermal conductivity in non-purified $^4$He below 120\,mK? As this happens near the same temperature, $T_0\sim80$\,mK at which the TO frequency shift appears in these samples, one might speculate that the same phenomenon might be responsible for both anomalies. 
%This is indirectly supported by the fact that in purified samples, for which $T_0 < 30$\,mK, no anomaly in $\kappa(T)$ was observed down to 80\,mK.
%Whatever this underlying phenomenon can be, it seems unlikely that it is superfluidity (that in liquid $^4$He boosts heat transport by means of the counterflow of the normal and superfluid components) -- because we have shown that a high-amplitude TO drive eliminates the TO frequency shift (the superfluid component would vanish at this point) while $\kappa$ stays unchanged.
%It would be tempting to attribute the observed plateau in thermal conductivity to the gas of $^3$He impuritons at concentration $x_3$ and m.\,f.\,p. of $d \sim 0.3$\,mm. However, even with a generous assumption of the impuriton band of $0.1$\,K, one can only account for $10^{-4}$ of the observed value of $\kappa \sim 0.5$\,Wm$^{-1}$K$^{-1}$. 

 To summarize, we found that in different samples of polycrystalline hcp $^4$He the magnitude of the TO frequency shift and the crystalline disorder as probed by thermal phonons are not correlated. The phonon m.\,f.\,p. readily increases upon annealing at $T=0.8T_{\rm m}$ while the value of the frequency shift hardly changes. Driving TO at high amplitude eliminates the TO frequency shift  but leaves thermal conductivity unchanged. 
 %The most unexpected result of the experiments was, however, the observation of a substantial excess thermal conductivity at low temperatures in samples of solid $^4$He with natural concentration of the $^3$He isotope. While appearing at the same temperatures as the TO anomaly, $T<120$\,mK, the enhanced thermal conductivity does not correlate with the mechanical behaviour: $\kappa$ was unchanged when the TO was driven at amplitudes high enough to eliminate the TO frequency shift. 
 All these observations allow to conclude that the defects responsible for the TO anomaly and for the phonon mean free path in hcp $^4$He have different properties. 
The discovered enhancement of the thermal conductivity at low temperatures 
%cannot be accounted by specular scattering of ballistic phonons at small angles of incidence; it 
is likely to be related to the recently observed excess contribution to the specific heat \cite{Chan2009} -- the origins of both are yet unknown and require further investigation. 

We thank H.\,J.~Maris for valuable comments, P.\,V.\,E.~McClintock for providing isotopically purified $^4$He, and R. Schanen and P. Mirthinti for their contribution during the early stages of the experiment. This work was supported by the
Engineering and Physical Sciences Research Council [grant number EP/H014691].


\begin{thebibliography}{99}

%\bibitem{Goodkind1997} P.-C. Ho, I.\,P. Bindloss and J.\,M. Goodkind, J. Low Temp. Phys. {\bf 109}, 409 (1997). 

\bibitem{Chan2004}
E.~Kim and M.\,H.\,W.~Chan, Science {\bf 305}, 1941 (2004).
%Nature (London) {\bf 427}, 225 (2004); 

\bibitem{Andreev69}
A.\,F.~Andreev and I.\,M. Lifshitz, Sov. Phys. JETP {\bf 29}, 1107 (1969).

%\bibitem{Chester70} G.\,V.~Chester, Phys. Rev. A {\bf 2}, 256 (1970).

\bibitem{Leggett70}
A.\,J.~Leggett, Phys. Rev. Lett. {\bf 25}, 1543 (1970).

%\bibitem{Reppy07} A.\,S.\,C. Rittner and J.\,D. Reppy, Phys. Rev. Lett. {\bf 98}, 175302 (2007).


\bibitem{Reppy06} A.\,S.\,C. Rittner and J.\,D. Reppy, Phys. Rev. Lett. {\bf 97}, 165301 (2006).

\bibitem{Chan2007}
A.\,C. Clark, J.\,T. West, M.\,H.\,W. Chan, Phys. Rev. Lett. {\bf 99}, 135302 (2007).

%\bibitem{ChanSlab} ChanSlab

%\bibitem{Chan2006} E. Kim and M.\,H.\,W. Chan, Phys. Rev. Lett. {\bf 97}, 115302 (2006).

\bibitem{Chan08}
E. Kim {\it et al.}, Phys. Rev. Lett. {\bf 100}, 065301 (2008).
%E. Kim, J.\,S. Xia, J.\,T. West {\it et al.}, Phys. Rev. Lett. {\bf 100}, 065301 (2008).

\bibitem{SuperfluidTheory} L. Pollet {\it et al.}, Phys. Rev. Lett. {\bf 101}, 097202 (2008).
%L. Pollet {\it et al.}, Phys. Rev. Lett. {\bf 98}, 135301 (2007). %grain boundaries
%dislocations: M. Boninsegni {\it et al.}, Phys. Rev. Lett. {\bf 99}, 035301 (2007).

\bibitem{Balatsky} 
J.-J. Su, M.\,J. Graf, and A.\,V.  Balatsky, Phys. Rev. Lett. {\bf 105}, 045302 (2010). 

\bibitem{Iwasa10} 
I. Iwasa, Phys. Rev. B {\bf 81}, 104527 (2010).

\bibitem{Armstrong79}
G.\,A.~Armstrong, A.\,A.~Helmy, and A.\,S.~Greenberg, Phys.~Rev.~B~{\bf 20}, 1061 (1979).

\bibitem{Ziman} J.\,M. Ziman, {\it Electrons and Phonons} (Oxford~U.~P., 1960).

%\bibitem{Mezhov66} L.\,P.~Mezhov-Deglin, Zh.~Eksp.~Teor.~Fiz. {\bf 49}, 66 (1965); Sov. Phys. JETP {\bf 22}, 47 (1966).

%\bibitem{Levchenko} A.\,A. Levchenko, L.\,P. Mezhov-Deglin, Lecture Notes in Physics {\bf 285}, 438 (1987). 
%A.\,A. Levchenko, L.\,P. Mezhov-Deglin, Zh. Eksp. Theor. Fiz. {\bf 86}, 2123 (1984); Sov. Phys. JETP {\bf 59}, 1234 (1984).

\bibitem{Chan2009} 
X. Lin, A.\,C. Clark, Z.\,G. Cheng, and M.\,H.\,W. Chan, Phys. Rev. Lett. {\bf 102}, 125302 (2009).

\bibitem{BeamishShearModulus} 
O. Syshchenko, J. Day, and J. Beamish, Phys. Rev. Lett. {\bf 104}, 195301 (2010). 
%\bibitem{Berman} Berman, Thermal Conductance ...

\bibitem{Maris1970} A.\,K. McCurdy, H.\,J. Maris, and C. Elbaum, Phys. Rev. B {\bf 2}, 4077 (1970).

\bibitem{Maris2011} H.\,J. Maris, private communication.

%\bibitem{Paalanen} M.\,A. Paalanen, D.\,J. Bishop, and H.\,W. Dail, Phys. Rev. Lett. {\bf 46}, 664 (1981). 

\bibitem{Hendry87}
P.\,C.~Hendry and P.\,V.\,E.~McClintock, Cryogenics {\bf 27}, 131 (1987).

\bibitem{Driessen86} A. Driessen, E. van der Poll, and I.\,F. Silvera, Phys. Rev. B {\bf 33}, 3269 (1986).
%E.\,R. Grilly and R.\,L. Mills, Ann. Phys. {\bf 8}, 1 (1959). 

\bibitem{Greywall77} D.\,S. Greywall, Phys. Rev. B {\bf 16}, 5127 (1977). 

%\bibitem{Chan06} E.~Kim and M.\,H.\,W.~Chan, Phys. Rev. Lett.~{\bf 97}, 115302 (2006).
%TO measurements, fitted by Iwasa2010

%\bibitem{Carruthers1961} P. Carruthers, Rev. Mod. Phys. {\bf 33}, 92 (1961). % slab d: Casimir scattering length L = (3\pi/8)d \approx 1.178d (pure boundary scattering - diffuse,  high aspect ratio slab - no length correction.) C_v=(\frac{4\pi^4k^4}{15h^3})(4\pi\sigma_\{\lambda}\frac{1}{c_{\lambda}^3})T^3; ${\bar v} \equiv \frac{\sigma c_\{\lambda}^{-2}}{\sigma c_\{\lambda}^{-3}}$


\bibitem{BurnsXray2008} C.\,A. Burns {\it et al.}, Phys. Rev. B {\bf 78}, 224305 (2008).

\bibitem{Eddison1985} C.\,G. Eddison and M.\,N. Wybourne, J. Phys. C {\bf 18}, 5225 (1985).

%\bibitem{Poiseuille} Poiseuille flow of phonon gas, known to result in $\kappa \propto T^6$ dependence \cite{Armstrong79}, cannot operate at $T<500$\,mK where phonon-phonon m.\,f.\,p. greatly exceeds the sample size.

%\bibitem{Rudavskii} V.\,N. Grigor'ev {\it et al.}, Phys. Rev. B {\bf 76}, 224524 (2007).

%\bibitem{Goodkind2002} J.\,M.~Goodkind, Phys. Rev. Lett. {\bf 89}, 095301 (2002).

\bibitem{Mezhov65} L.\,P. Mezhov-Deglin, Zh. Eksp. Theor. Fiz. {\bf 49}, 66 (1965).

%\bibitem{Reppy10} J.\,D. Reppy, Phys. Rev. Lett. {\bf 104}, 255301 (2010).

%\bibitem{Mezhov65} Mezhov65

%\bibitem{Goodkind93} C.\,A.~Burns and J.\,M.~Goodkind, J. Low Temp. Phys.~{\bf 93}, 15 (1993).
%thermal conductivity 80-500mK, perfect T^3

%\bibitem{counterflow} In superfluid $^4$He driven at low velocities, the rotational inertia is reduced due to the decoupling of the superfluid component from TO oscillations, and the heat transfer is enhanced via the counterflow of the normal and superfluid components. However, after exceeding the critical velocity the superfluid component vanishes, and so do the deficit of inertia and the anomalous heat transfer.
%This should not be confused with the enhanced heat transfer in liquid $^4$He below $T_{\lambda}$ carried by the internal convection of the normal component (fluid of thermal excitations) counterflowing with the ordered superfluid component; in the presence of the special reference frame associated with solid helium internal convection of the normal component relative to the solid becomes impossible. 

\end{thebibliography}
\end{document}